\documentclass[conference]{IEEEtran}
\IEEEoverridecommandlockouts
\usepackage{cite}
\usepackage{url}
\usepackage{hyperref}
\usepackage{amsmath,amssymb,amsfonts}
\usepackage{algorithm}
\usepackage{algorithmic}
\usepackage{graphicx}
\usepackage{booktabs}
\usepackage{multirow}
\usepackage{tcolorbox}
\tcbuselibrary{skins,breakable,xparse}
\usepackage{array} 
\usepackage{textcomp}
\usepackage{amsmath}
\usepackage{fancyvrb} 
\usepackage{fvextra}  
\usepackage{caption}  
\usepackage{xcolor}
\def\BibTeX{{\rm B\kern-.05em{\sc i\kern-.025em b}\kern-.08em
    T\kern-.1667em\lower.7ex\hbox{E}\kern-.125emX}}

\DefineVerbatimEnvironment{json}{Verbatim}{
    fontsize=\small,       
    baselinestretch=0.9,   
    breaklines=true,       
    breakanywhere=true,    
    numbers=left,          
    numbersep=5pt,         
    frame=none,            
    xleftmargin=15pt,      
}

\begin{document}

\title{LogSage: An LLM-Based Framework for CI/CD Failure Detection and Remediation with Industrial Validation}

\author{
\IEEEauthorblockN{
    Weiyuan Xu\textsuperscript{1,}\textsuperscript{2,}\textsuperscript{†},
    Juntao Luo\textsuperscript{2,}\textsuperscript{†,}\textsuperscript{‡},
    Tao Huang\textsuperscript{2},
    Kaixin Sui\textsuperscript{2},
    Jie Geng\textsuperscript{2},\\
    Qijun Ma\textsuperscript{2},
    Isami Akasaka\textsuperscript{2},
    Xiaoxue Shi\textsuperscript{2},
    Jing Tang\textsuperscript{2},
    Peng Cai\textsuperscript{1,}\textsuperscript{*}
}
\IEEEauthorblockA{\textsuperscript{1}East China Normal University, Shanghai, China}
\IEEEauthorblockA{\textsuperscript{2}ByteDance, Shanghai, China}
\IEEEauthorblockA{
}
\thanks{\textsuperscript{*}\;Corresponding author.\textsuperscript{†}\;Equal contribution.\textsuperscript{‡}\;Work conducted during the internship at ByteDance.}
}

\maketitle

\begin{abstract}
Continuous Integration and Deployment (CI/CD) pipelines are critical to modern software engineering, yet diagnosing and resolving their failures remains complex and labor-intensive. We present LogSage, the first end-to-end LLM-powered framework for root cause analysis (RCA) and automated remediation of CI/CD failures. LogSage employs a token-efficient log preprocessing pipeline to filter noise and extract critical errors, then performs structured diagnostic prompting for accurate RCA. For solution generation, it leverages retrieval-augmented generation (RAG) to reuse historical fixes and invokes automation fixes via LLM tool-calling. 

On a newly curated benchmark of 367 GitHub CI/CD failures, LogSage achieves over 98\% precision, near-perfect recall, and an F1 improvement of more than 38\% points in the RCA stage, compared with recent LLM-based baselines. In a year-long industrial deployment at ByteDance, it processed over 1.07M executions, with end-to-end precision exceeding 80\%. These results demonstrate that LogSage provides a scalable and practical solution for automating CI/CD failure management in real-world DevOps workflows.
\end{abstract}


\section{Introduction}

As the backbone of modern software engineering, Continuous Integration and Continuous Deployment (CI/CD) has become critical infrastructure for reliable, rapid software delivery \cite{Fitzgerald2017}. It enables higher release frequency, faster iteration, and reduced operational risk, and is widely adopted across leading technology companies \cite{microsoft, meta, github}. Large-scale empirical studies also confirm its growing adoption in open-source communities and its pivotal role in modern development practices \cite{hilton2016ci}.

In parallel, large language models (LLMs) have achieved impressive results across a range of software engineering tasks \cite{hou2024largelanguagemodels}, including requirements engineering \cite{hemmat2025llm}, code retrieval \cite{li2024reco}, automated code review \cite{bitscr}, and unit test enhancement \cite{Alshahwan2024}. Encouraged by these advances, recent work has begun exploring LLM-based approaches for failure detection and remediation—opening up new possibilities for intelligent automation in software delivery.

Yet despite progress in automated software engineering, real-world CI/CD pipelines still suffer frequent failures. Diagnosing and fixing them typically requires on-call engineers to inspect lengthy, semi-structured logs filled with irrelevant or misleading information, a process that is time-consuming and detrimental to both development velocity and product stability. While many failures follow recurring patterns and organizations accumulate substantial resolution knowledge, this knowledge is often stored in unstructured formats that are difficult to retrieve or reuse with traditional non-LLM methods, leading to redundant work and wasted resources.

A major barrier is the limited academic focus on CI/CD log analysis. Although log-based anomaly detection is well studied, most work targets streaming system logs \cite{howfar} and assumes structured formats or consistent templates \cite{neuralog,deeplog,loganomaly,semi-supervised,logrobust}, which generalize poorly to noisy, context-rich, file-level CI/CD logs. Even recent CI/CD log analysis efforts \cite{uniloc2023} rely on template-based deep learning techniques with poor generalizability. These approaches often require large labeled datasets for training, lack interpretability, and cannot produce actionable remediation strategies. While LLMs offer promising reasoning capacity, their application to CI/CD root cause analysis remains underexplored \cite{10586953}, and existing baselines leave considerable room for improvement in diagnostic accuracy, repair guidance, and end-to-end integration.

To address these gaps, we present \textbf{LogSage}, the first end-to-end LLM-powered framework for CI/CD failure detection and remediation. LogSage conducts fine-grained root cause analysis (RCA) and automated resolution directly from raw logs, producing human-readable diagnostic reports and applying executable fixes through tool-calling. Its two-stage pipeline combines token-efficient log preprocessing for RCA with multi-route retrieval-augmented generation (RAG) for solution generation, retrieving domain knowledge from diverse internal sources and integrating it with RCA report to synthesize executable remediation strategies. This enables accurate RCA, interpretable reasoning, and automated recovery with minimal developer effort. Our key contributions are as follows:

\begin{itemize}
    \item \textbf{First end-to-end LLM framework:} LogSage is the first LLM-based framework for fully automated CI/CD failure detection and remediation. It integrates a token-efficient log preprocessing strategy that filters noise, expands context, and adheres to token constraints without relying on templates, together with a multi-route RAG-based solution generator that reuses historical fixes and leverages LLM tool-calling for executable remediation.
    \item \textbf{Curated dataset:} We release a dataset of \textbf{367} real-world CI/CD failures from GitHub repositories, each annotated with key log evidence, enabling reproducible RCA evaluation and fostering future research (see Appendix \ref{appendix}).
    \item \textbf{Extensive validation:} On the curated dataset, LogSage improves RCA F1 score by over \textbf{38\%} compared with prompt-based LLM baselines, achieving \textbf{98\% precision} and near-perfect recall. In large-scale industrial deployment at ByteDance, it processed over \textbf{1.07M executions} with sustained adoption and \textbf{80\%+ end-to-end precision}, demonstrating practical usability in production.
\end{itemize}

The remainder of this paper is structured as follows: Section~\ref{sec:relatedwork} surveys related work. Section~\ref{sec:method} introduces the LogSage framework, including both its high-level architecture and implementation details. Section~\ref{sec:experimental} presents experiments and evaluates the RCA stage. Section~\ref{sec:industrial} reports on the real-world deployment of LogSage at ByteDance. Finally, Section~\ref{sec:conclusion} concludes the paper and outlines directions for future research.

\begin{figure*}[htbp]
    \centering
    \includegraphics[width=1.0\textwidth]{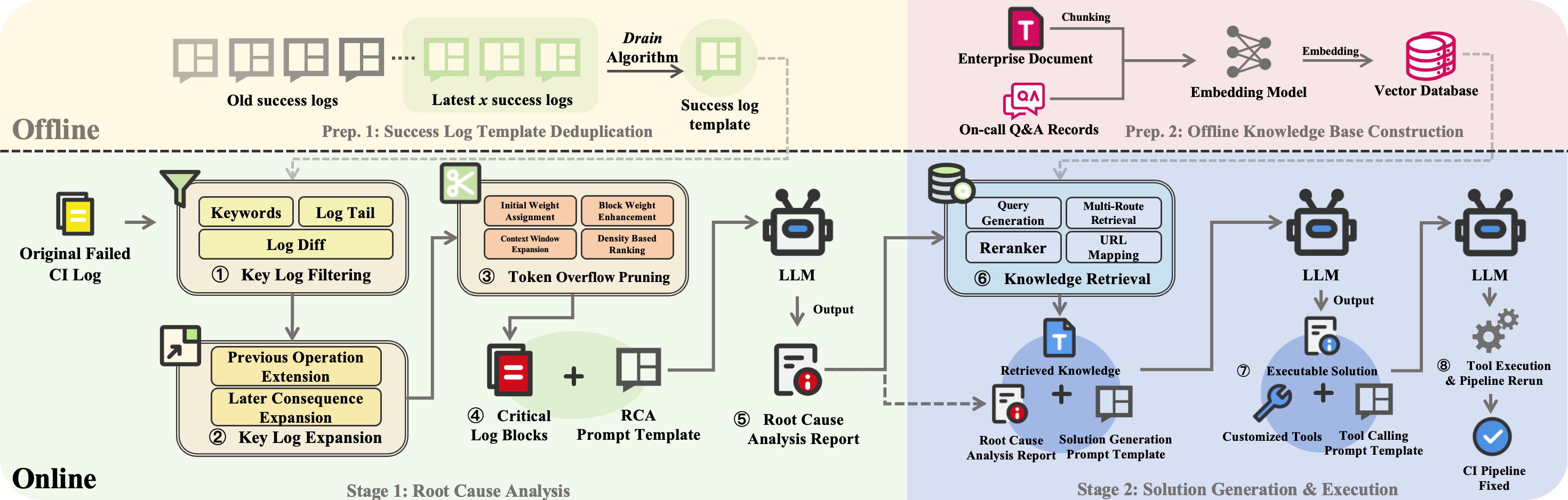}
    \caption{
        Overview of the \textbf{LogSage framework}, consisting of an offline preparation phase for log template deduplication and knowledge base construction, and an online operational phase for RCA and solution generation with execution.
    }
    \label{framework}
\end{figure*}

\section{Related Work}\label{sec:relatedwork}

With the rise of AI techniques in software engineering, CI/CD pipelines have increasingly adopted machine learning (ML), deep learning (DL), and LLMs to improve efficiency, reliability, and fault tolerance \cite{Mohammed2024}. In this section, we review prior work from three perspectives: traditional AI methods in CI/CD pipelines, log anomaly detection, and the emerging application of LLMs for failure detection and remediation.

\subsection{AI in CI/CD Pipelines}

Traditional ML approaches in CI/CD focus primarily on predicting test outcomes, build success, or optimizing test execution to reduce cost. These methods typically rely on structured features such as code metrics and commit history, with limited support for runtime failure analysis or recovery. For instance, CNNs have been used to identify false positives in static code analysis \cite{Lee2019}, while other studies aim to skip unnecessary tests or prioritize test selection to save time and cost \cite{Grano2018, Grano2019, AlSabbagh2019b, AlSabbagh2019, Abdalkareem2021}. Additionally, defect prediction has been explored as an indirect way to reduce failures, though these methods do not directly address fault localization or remediation \cite{Koroglu2016, Mamata2022}.

DL methods enhance predictive accuracy and generalization in failure prediction tasks \cite{Mishra2024, Saidani2022}. However, most models lack semantic understanding of failure causes. Mahindru et al.\cite{Mahindru2021} proposed the LA2R system, which combines log anomaly detection with metadata prediction to retrieve relevant resolutions. While effective, this approach depends heavily on templates and rule-based clustering, limiting its adaptability to evolving log formats and unseen failures.

\subsection{Log Anomaly Detection}

Among CI/CD tasks, log anomaly detection is particularly critical and challenging due to the unstructured and context-dependent nature of log data. Early deep learning work by Du et al.\cite{deeplog} used LSTMs to learn normal log sequences and detect anomalies. Meng et al.\cite{loganomaly} extended this idea with Template2Vec to capture sequential and quantitative anomalies. Yang et al.\cite{semi-supervised} employed a GRU-based attention model and semi-supervised learning to reduce labeling cost, while Zhang et al.\cite{logrobust} addressed unstable logs using attention-based Bi-LSTMs. Recent work by Le et al.\cite{howfar, neuralog} introduced parser-independent Transformer-based methods that improve generalization by avoiding reliance on log parsing.

Given their strong semantic understanding and reasoning abilities, LLMs are well suited for log anomaly detection. Studies have shown that even prompt-based LLMs can detect anomalies in streaming logs \cite{Xiao2024, Egersdoerfer2023, interpreteable, Liu2024}. Qi et al.\cite{loggpt} proposed a GPT-4-based detection framework, while Shan et al.\cite{faceit} leveraged configuration knowledge for anomaly detection. Almodovar et al.\cite{Almodovar2024} fine-tuned LLMs for better performance, and Ju et al.\cite{Ju2024} adopted RAG to enhance log interpretation. Hybrid methods have also been proposed: Guan et al.\cite{LogLLM} combined BERT and LLaMA for semantic vector extraction, and Hadadi et al.\cite{LLM_ML} used LLMs to compensate for missing context in unstable environments. In-context learning techniques were further used to refine log representation and guide fine-tuning \cite{Fariha2024, He2024}.

Despite these advances, most studies focus on streaming logs, with limited attention to file-based CI/CD logs. Moreover, few works provide large-scale industrial validation or full pipeline integration, leaving a gap for further exploration.

\subsection{LLMs for Failure Detection and Remediation}

Beyond anomaly detection, LLMs offer unique potential in end-to-end failure diagnosis and automated remediation. A recent survey identified root cause analysis and remediation as key use cases for LLMs in AIOps \cite{aiops_survey}. For root cause analysis, Roy et al.\cite{RCA} and Wang et al.\cite{Wang2024} proposed tool-augmented LLM agents, while Zhang et al.\cite{cloud_gpt4} and Goel et al.\cite{Goel2024} used in-context learning and RAG for incident diagnosis with real-world data. On the remediation side, Sarda et al.\cite{Sarda2024, Sarda2023} and Ahmed et al.\cite{rc_cloud_incidents} automated fault recovery in cloud environments. Wang et al.\cite{LogExpert} incorporated Stack Overflow into LLM-based solution generation, and Khlaisamniang et al.\cite{Khlaisamniang2023} applied generative AI for self-healing systems.

In the CI/CD context, however, LLM-based solutions remain underexplored. Chaudhary et al.\cite{Chaudhary2024} proposed an LLM-based assistant for CI/CD operations, and Sharma et al.\cite{Sharma2024} outlined key challenges in adopting LLMs for fault handling in build pipelines. While promising, these efforts have yet to address the complexity of real-world CI/CD logs or provide comprehensive remediation workflows.

Overall, while ML and DL methods have laid the foundation for predictive CI/CD analytics, LLMs bring new opportunities for building robust, context-aware, and automated failure detection and remediation systems. However, systematic integration of LLMs into CI/CD pipelines, especially for file-based logs and knowledge-driven automated remediation, remains an open research challenge.

\section{Methodology}\label{sec:method}

In this section, we first provide an overview of the LogSage framework architecture in Section~\ref{subsec:overview}, then we elaborate on the detailed workflows and key components of the two stages of LogSage in Section~\ref{subsec:rca} and Section~\ref{subsec:solutiongeneration}, respectively.

\subsection{Overview of LogSage}\label{subsec:overview}

The two-stage architecture of LogSage is illustrated in Figure~\ref{framework}. The system consists of both offline and online phases. In the offline phase, success log templates are deduplicated via the Drain algorithm (Prep.~1). Enterprise documents and on-call Q\&A records are embedded into vector databases to construct a knowledge base (Prep.~2). 
In the online phase, Stage~1 processes failed logs through (1) key log filtering, (2) expansion, and (3) token pruning, 
yielding (4) critical log blocks that form RCA prompt to generate (5) RCA report. 
Stage~2 leverages (6) knowledge retrieval and LLM-based tool selection to generate (7) executable solutions, followed by (8) automated tool execution and pipeline rerun, ultimately restoring the CI/CD pipeline.

\subsection{Root Cause Analysis Stage}\label{subsec:rca}
In this section, we present the design and implementation of the RCA stage. The goal of this stage is to extract the most relevant portions of raw CI/CD failure logs and enable accurate reasoning by the LLM, which then returns a structured diagnostic report containing key error lines and an interpretable root cause summary. This stage must address several practical challenges inherent in real-world CI/CD log analysis:

First, \textbf{log heterogeneity}: CI/CD pipelines differ widely in their execution steps, commands, output formats, and logging styles, making it infeasible to rely on a unified, structured template to accommodate all scenarios.

Second, \textbf{informational noise and misleading lines}: Raw logs often contain numerous irrelevant \texttt{WARNING} or \texttt{ERROR} messages that do not reflect the actual cause of failure. Naively feeding the full log into the LLM can lead to degraded reasoning quality and even hallucinations. Conversely, filtering logs too aggressively (e.g., by extracting only lines containing keywords such as \texttt{ERROR} or \texttt{FAILED}) may omit critical contextual information, causing the model to speculate and increasing the risk of misdiagnosis.

Third, \textbf{input length constraints}: LLMs cannot accept arbitrarily long inputs. Excessively long log sequences not only increase inference cost and latency but also reduce reasoning accuracy due to context dilution. In industrial CI/CD environments, these limitations become particularly critical, as practical deployments require LLM applications to maintain a stable context length and predictable response latency in order to ensure system robustness and operational usability.

To address these challenges, we design a dedicated log preprocessing pipeline prior to model invocation. This pipeline consists of the following modules:

\begin{itemize}
    \item \textbf{Key Log Filtering}: Extracts candidate log lines from the \textit{failed log} based on keyword matching, log tail prioritization and log diff against recent \textit{success log template}.
    \item \textbf{Key Log Expansion}: Adds contextual lines surrounding the extracted errors to preserve semantic coherence and prevent information loss.
    \item \textbf{Token Overflow Pruning}: Ranks expanded log blocks by weight calculation and prunes low-priority blocks to ensure input remains within a predefined token limit.
    \item \textbf{Dynamic Prompt Assembly}: Constructs RCA prompt by integrating role-playing, chain-of-thought reasoning, few-shot learning, and output-format constraints. Then dynamically combines them with the processed critical log blocks to guide the LLM toward accurate and reproducible RCA reasoning.
\end{itemize}

The processed critical log blocks are delivered to the LLM through a structured prompt, allowing the model to perform accurate reasoning and generate a root cause analysis report.

\vspace{0.5em}
\subsubsection{\textbf{Key Log Filtering Module}}

This module is designed for accurately extracting relevant error log lines from the \textit{failed log} by combining \textit{log diff, keyword matching} and \textit{log tail prioritization}. These strategies collectively ensure broad and precise coverage of failure-relevant log lines. The filtering process is formalized in Algorithm \ref{alg:log-filtering}.

\begin{algorithm}[ht]
\caption{Key Log Filtering Process}
\label{alg:log-filtering}
\begin{algorithmic}[1]
\STATE \textbf{Input:} $failed\_log$, $success\_log\_templates$ \COMMENT{from offline preparation}
\STATE \textbf{Output:} $filtered\_log$
\STATE Initialize $candidate\_pool \leftarrow \{\}$
\FOR{each line $l$ in $failed\_log$}
    \STATE $template \leftarrow \mathrm{extract\_template}(l)$
    \STATE $position \leftarrow \mathrm{get\_position}(l, failed\_log)$
    \IF{$template \notin success\_log\_templates$}
        \STATE Add $l$ to $candidate\_pool$ \COMMENT{log diff}
    \ENDIF
    \IF{$\mathrm{contains\_keyword}(l)$}
        \STATE Add $l$ to $candidate\_pool$ \COMMENT{keyword matching}
    \ENDIF
    \IF{$\mathrm{is\_in\_log\_tail}(position)$}
        \STATE Add $l$ to $candidate\_pool$ \COMMENT{log tail prioritization}
    \ENDIF
\ENDFOR
\STATE $filtered\_log \leftarrow \mathrm{deduplicate}(candidate\_pool)$
\RETURN $filtered\_log$
\end{algorithmic}
\end{algorithm}

\textit{\textbf{Log diff}} is the most critical strategy in the key log filtering module. It leverages the repetitive nature of CI/CD pipelines: executions typically share almost identical configurations across consecutive runs, yielding highly consistent log outputs during successful executions, and even when changes occur, they are usually incremental and persist across multiple subsequent runs. Exploiting this stability, the system performs an offline process that applies the \textit{Drain} algorithm~\cite{He2017} to recent \textit{success logs} of the same pipeline, extracting structural templates that characterize stable and recurring log lines. These templates are then stored and used online to filter the \textit{failed log}: lines matching the success-run templates are treated as background outputs that are highly unlikely to contain failure-related information, and are therefore excluded from the error candidate set. This log-diff approach effectively eliminates misleading \texttt{WARNING} or \texttt{ERROR} lines while avoiding rigid handcrafted rules, thereby significantly reducing noise in downstream analysis.

To maintain the freshness and relevance of filtering templates derived from \textit{success logs}, LogSage adopts an offline \textit{\textbf{log template deduplication}} strategy: for each pipeline task, only the most recent $x$ successful logs are retained for template extraction, where $x$ is configurable. Empirical analysis across diverse CI/CD projects in ByteDance shows that setting $x=3$ achieves the best trade-off between template diversity and noise reduction, and is used as the default in our system. The value of $x$ can be tuned based on pipeline stability and log variance, thereby improving the generalizability of LogSage.

\vspace{0.5em}
\textit{\textbf{Keyword matching}} identifies log lines containing high-risk terms based on a curated set of failure-related keywords mined from historical CI/CD failure cases. Any log line matching one or more of these keywords is added to a candidate pool for further downstream processing. The keyword set includes:

\begin{quote}
\texttt{fatal, fail, panic, error, exit, kill, no such file, err:, err!, failures:, err , missing, exception, cannot}
\end{quote}

\vspace{0.5em}
The \textit{\textbf{log tail prioritization}} strategy is motivated by the empirical observation: most critical error logs tend to appear near the end of the file, as CI/CD pipeline failures often lead to abrupt termination. Accordingly, log lines appearing at the end of the \textit{failed log} are prioritized during candidate selection.

\vspace{0.5em}
Through these strategies, the \textit{Key Log Filtering} module is able to deconstruct the \textit{failed log} with high precision and identify all potentially problematic log lines. After removing redundant lines, the filtered logs are structured as pairs of line numbers and corresponding log lines, serving as input for subsequent processing.

\vspace{0.5em}
\subsubsection{\textbf{Key Log Expansion Module}}

Based on manual experience in analyzing CI/CD failure logs, it is often insufficient to rely solely on the \texttt{ERROR} log lines that directly report failures. The true root cause of a pipeline error is typically identified by examining several lines before and after the failure line. This observation motivated the design of the \textit{Key Log Expansion} module, which provides additional contextual information to support LLM-based root cause analysis, helping to mitigate hallucinations and logical errors caused by missing context.

Starting from the key error lines identified by the previous module, the expansion module includes $m$ lines before and $n$ lines after each key line to form a log block. We adopt an asymmetric expansion strategy where $n > m$: the preceding $m$ lines ensure that the operations leading to the error are captured, while the succeeding $n$ lines provide richer post-error information such as stack traces to cover information that is often more critical than the pre-error context. Overlapping blocks resulting from multiple key lines are merged into a single cohesive block to maintain contextual continuity. This expansion ensures that the LLM can access sufficient surrounding information to interpret the key log lines accurately.

In practice, we set $m = 4$ and $n = 6$ for both online deployments and offline experiments. Empirical results show that this configuration is sufficient to retain most of the contextual information necessary for accurate root cause analysis.

\vspace{0.5em}
\subsubsection{\textbf{Token Overflow Pruning Module}}
In practical production environments, the token length limitation of LLMs imposes a critical constraint, as overly long input tokens can reduce compliance with instructions, introduce hallucinations, and increase computational cost and latency. To address this, LogSage incorporates a \textit{Token Overflow Pruning} module that enforces a predefined token limit during RCA. This module leverages a structured log weighting and enhancement algorithm to assess the relative importance of expanded log blocks produced in the previous module. This module includes four components: \textit{initial weight assignment} to identify candidate lines, \textit{pattern-based weight enhancement} to emphasize critical error lines, \textit{contextual window expansion} to preserve semantic continuity, and \textit{density-based block ranking} to prioritize the most informative blocks while satisfying the token constraint.

\vspace{0.5em}
\textit{\textbf{Initial Weight Assignment:}} Given the \textit{failed log} $L = \{l_1, l_2, ..., l_n\}$ and log lines from the candidate pool $I = \{i_1, i_2, ..., i_m\}$, we define a weight list $W = \{w_1, w_2, ..., w_n\}$, where each $w_j$ represents the weight assigned to line $l_j$:

\begin{equation}
w_j =
\begin{cases}
3 & \text{if } \frac{|I|}{|L|} \leq \alpha  \text{ and } |I| \leq \beta \text{ and } j \in \text{candidate pool}\\
1 & \text{if } \; j \in \text{candidate pool} \\
0 & \text{otherwise}
\end{cases}
\label{equa:predefine}
\end{equation}

This adaptive rule (Equation \ref{equa:predefine}) adds an initial weight to each log in the candidate pool and ensures that higher weights are assigned to sparse yet informative candidate log lines. The parameters $\alpha$ and $\beta$ are tunable thresholds that control the criteria for assigning high weights. Based on empirical observations, we set $\alpha = 0.7$ and $\beta = 500$ in practice.

\vspace{0.5em}
\textit{\textbf{Pattern-Based Weight Enhancement.}} To emphasize critical log lines, we apply a rule-based scheme: (i) lines containing typical failure markers (e.g., \texttt{--- FAIL:}, \texttt{Failures:}) are assigned the maximum weight $w_i = 10$; (ii) lines with curated keywords or section headers (e.g., starting with \texttt{\#}) receive a moderate boost ($w_i = w_i + 2$); and (iii) remaining candidates in the recall pool are lightly reinforced with $w_i = w_i + 1$. This design ensures both explicit and implicit failure signals are prioritized in subsequent pruning and diagnosis.

\vspace{0.5em}
\textit{\textbf{Contextual Window Expansion:}} To ensure the contextual integrity of high-weight critical log lines, log lines with weights above the threshold $\theta$ are expanded into log blocks. For any $w_i \geq \theta$, the log blocks in its neighborhood $[i - m, i + n]$ are added to the candidate pool, where $m$ and $n$ represent the number of previous and next lines as defined previously.

The threshold $\theta$ is adaptively defined by Equation \ref{equa: theta} according to two different situations:

\begin{equation}
\theta =
\begin{cases}
1 & \text{if} \; \max(W) = 1 \text{ or } |\{w_i \geq 1\}| \leq \gamma \\
3 & \text{otherwise}
\end{cases}
\label{equa: theta}
\end{equation}

In the first scenario, all key log lines receive a uniform weight of 1, or the number of high-weight lines falls below a threshold $\gamma$, suggesting that the preceding filtering step has limited effectiveness in isolating truly critical lines. In this case, broader contextual expansion is required to ensure the LLM has sufficient information. In contrast, when high-weight lines are adequately identified, context expansion is applied selectively around those lines to maintain focus and efficiency. In practice, we empirically set $\gamma = 500$.

\vspace{0.5em}
\textit{\textbf{Density-Based Block Ranking:}} In order to compare the weights between different log blocks, we define the log block weight density. For contiguous candidate log lines that are grouped into blocks $B_j = [s_j, e_j]$, their weight density is computed as Equation \ref{equa: density}:

\begin{equation}
\text{density}(B_j) = \frac{\sum_{i=s_j}^{e_j} w_i}{e_j - s_j + 1}
\label{equa: density}
\end{equation}

After computing the weight density for all candidate log blocks, they are ranked in descending order of density and indexed as $B_1, B_2, \dots, B_K$, where $k$ denotes the rank of the $k$-th block in this sorted list. A greedy selection algorithm is then applied to include as many blocks as possible while ensuring that the total token count does not exceed the predefined limit:
\begin{align*}
\text{Token Limit} \geq \sum_{i=1}^{k-1} \text{Token}(B_i) + \text{Token}(B_k)
\end{align*}

Based on empirical observations, we set the token limit at 22,000, which provides sufficient contextual coverage while maintaining a reasonable computational cost. Under this constraint, high-density log blocks are retained with priority, while low-density blocks exceeding the token limit are discarded to achieve effective pruning.

\vspace{0.5em}
Overall, this weighted pruning strategy strikes a practical balance between capturing critical error information and controlling input length. It is well-defined and tunable, allowing parameter adjustments across diverse real-world CI/CD systems, and thereby demonstrating strong generalizability.

\vspace{0.5em}
\subsubsection{\textbf{Root Cause Analysis Prompt Template Design}}

To facilitate RCA, LogSage employs a task-specific prompt template (Fig. \ref{rca_prompt}) that incorporates prompt engineering techniques such as role-playing, few-shot learning, and output format constraints. This template provides the LLM with explicit task instructions and takes filtered critical error log blocks as input, guiding the model to focus on diagnostic reasoning.

\subsection{Solution Generation and Execution Stage}\label{subsec:solutiongeneration}

In this section, we present the design and implementation of the solution generation and execution stage. LogSage combines critical log blocks and RCA report from the previous stage with domain-specific knowledge retrieved via a multi-route RAG mechanism. The system constructs a prompt enriched with RCA report, domain knowledge and well-designed tools, enabling the LLM to produce executable remediation suggestions and automatically select and invoke appropriate repair tools to resolve CI/CD pipeline failures.

\subsubsection{\textbf{Offline Knowledge Base Construction}}

At ByteDance, a large volume of domain knowledge has been accumulated through CI/CD platform operations. This includes:

\begin{itemize}
    \item 1,206 Feishu documents detailing production CI/CD issues and resolutions, contributed by various teams.
    \item 23,344 historical on-call Q\&A pairs recorded by rotating engineering teams responsible for incident response.
\end{itemize}

Feishu documents are segmented using a chunking strategy with a 3,000 max-token cutoff and embedded via LLM-based encoders. Q\&A pairs, due to their brevity, are stored directly as \texttt{<question, answer>} entries. To ensure high availability, the entire knowledge base is replicated across both \textit{VikingDB} and \textit{Elasticsearch}, enabling failover retrieval in production environments.

\subsubsection{\textbf{Online Multi-Route Retrieval Mechanism}}
Online process consists of four steps: \textit{query construction}, \textit{coarse retrieval}, \textit{reranking}, and \textit{}{URL mapping}.

\vspace{0.5em}
\textit{\textbf{RAG Query Construction:}} To mitigate the inherent variability in natural-language root cause descriptions, we reformulate the RAG query by combining the LLM-generated root cause with the corresponding critical log snippet. This hybrid query construction enhances retrieval efficiency and accuracy by better aligning contextual semantics. For example:

\begin{tcolorbox}[colback=gray!3, colframe=gray!60!black, boxrule=0.3pt, arc=2pt, left=4pt, right=4pt, top=4pt, bottom=4pt]
\scriptsize
\textbf{LLM Output:} \texttt{Unit test \textnormal{TestFilterPushCdnOnCreate} failed.} \\[2pt]
\textbf{\textcolor{gray!60!black}{CI Error Message:}} \\[-2pt]
\begin{quote}
\ttfamily
Warn 2024-04-26 18:08:39,457 v1(7) stream\_create.go:1524 ... unmarshal err ... ReadMapCB: expect \{ or n, but found \textbackslash x00 ...
\end{quote}
\end{tcolorbox}

\textit{\textbf{Multi-Route Coarse Retrieval:}} LogSage's coarse-grained retrieval combines three orthogonal dimensions: (1) \emph{database infrastructure}, including VikingDB and Elasticsearch; (2) \emph{matching granularity}, using both \texttt{query2doc} (query-to-content) and \texttt{query2query} (query-to-title) strategies; and (3) \emph{similarity metric}, supporting \texttt{BM25} (sparse lexical) and \texttt{KNN} (dense vector) retrievals. These combinations yield six base retrieval paths. To further enhance recall coverage, we integrate two auxiliary routes: \texttt{lark\_wiki} (for enterprise documents via Lark search API) and \texttt{rds\_match} (for internal historical query log mining). Thus resulting in 8 coarse retrieval routes:

\begin{tcolorbox}[colback=gray!2!white, colframe=gray!50!black, boxrule=0.3pt, arc=2pt, top=2pt, bottom=2pt]
\scriptsize
\begin{center}
\begin{tabular}{@{}ll@{}}
\texttt{query2doc\_viking\_knn} & \texttt{query2query\_viking\_knn} \\
\texttt{query2doc\_es\_knn}     & \texttt{query2doc\_es\_keyword} \\
\texttt{query2query\_es\_keyword} & \texttt{query2query\_es\_knn} \\
\texttt{lark\_wiki}             & \texttt{rds\_match}
\end{tabular}
\end{center}
\end{tcolorbox}

\textit{\textbf{Reranking:}} After coarse retrieval, reranking is performed using BGE (BAAI General Embedding) model and cosine similarity:

\begin{itemize}
    \item For Feishu documents, the entire chunk is used to compute similarity, as titles are often unaligned with specific CI/CD issues. The top 10 results are retained.
    \item For Q\&A pairs, since the query already aligns with the ``question'' field, only the top 100 coarse candidates are reranked, and the top 10 are selected.
\end{itemize}

\vspace{0.5em}
\textit{\textbf{URL Mapping:}} To prevent hallucinations caused by long URLs during generation, all retrieved links are replaced with numbered placeholders (e.g., \texttt{[CI Guide](files\_0)}) before being passed to the LLM. Once a solution is generated, the placeholders are mapped back to the original URLs, allowing users to access the full content for further reference.

\begin{figure}[htbp]
    \centering
\begin{tcolorbox}[title=\textbf{Root Cause Analysis Prompt Template}, colback=gray!5!white, colframe=gray!75!black, fonttitle=\bfseries, boxrule=0.5pt, arc=2pt, left=1pt, right=1pt, top=4pt, bottom=4pt]
    \scriptsize

\textbf{\# Role:} You are a CI/CD failure diagnosis assistant. Your task is to identify the root cause of pipeline failures based on execution logs and configuration info. \\

\textbf{\# Skills:}
\begin{itemize}
    \item \textbf{Task Type Identification:} Read config files to determine the task type (e.g., unit test, code scan). Output under \texttt{Diagnosis Process → Task Type}.
    \item \textbf{Error Log Analysis:} Read logs to identify up to 10 key error lines. Focus on terminal and causal errors. Output as \texttt{line range + conclusion}. Do NOT analyze normal/warning logs.
    \item \textbf{Root Cause Inference:} Use log and config analysis (don't mix unrelated errors). List up to 3 likely causes with concrete names and detailed, objective explanation. No fix suggestions.
\end{itemize}

\textbf{\# Output Format:}
\begin{itemize}
    \item Two parts: \texttt{Diagnosis Process} and \texttt{Root Cause}.
    \item For \texttt{Diagnosis Process}, include:
    \begin{itemize}
        \item Task type: e.g., \texttt{Run npm dependency installation}
        \item Error analysis: e.g., \texttt{Lines 6–12: Unit test `abc` failed due to result mismatch}
        \item Summarize causally related/similar errors in one line
        \item When referencing too many lines, use only first 5 + \texttt{etc.}
    \end{itemize}
    \item For \texttt{Root Cause}, format each cause as:
\begin{verbatim}
[High Likelihood] Unit test `abc` failed due to ...
\end{verbatim}
    \item Prefer one cause, max three. Use concrete info (test name, file, dep).
\end{itemize}

\textbf{\# Notes:}
\begin{itemize}
    \item Be concise and factual. Use "lines a, b, c–d" format when needed.
    \item Use \texttt{inline code} for log lines, \texttt{code blocks} for log content.
    \item No fix suggestions allowed.
    \item All results will be used for solution generation. Follow rules strictly.
\end{itemize}

\textbf{\# Constraints:}
\begin{itemize}
    \item DO NOT include normal/process/non-critical logs.
    \item DO NOT analyze similar/adjacent logs separately.
    \item DO NOT output more than 5 log line references without using \texttt{etc.}
\end{itemize}

\end{tcolorbox}
\caption{Prompt template for LogSage’s root cause analysis stage.}
    \label{rca_prompt}
\end{figure}

\subsubsection{\textbf{Solution Generation and Automated Execution}}

Upon completing knowledge retrieval, LogSage proceeds to the remediation stage, where it synthesizes solutions based on RCA report and retrieved domain knowledge. This stage explicitly decouples \textit{solution generation} from \textit{tool execution}, improving output focus by separating reasoning-oriented and action-oriented prompts.

LogSage prioritizes generating high-quality, actionable suggestions grounded in enterprise-specific knowledge. While internal tools can automate a subset of typical failures, many CI/CD issues remain beyond full automation. For such cases, LogSage still provides structured and executable guidance for developers, balancing intelligent assistance with interpretability rather than pursuing full end-to-end autonomy.

\vspace{0.5em}
\textit{\textbf{Prompt Construction for Solution Generation:}} To generate high-quality remediation suggestions, the system populates a predefined prompt template (Fig. \ref{suggestion_prompt}) with the RCA report, critical log blocks, and retrieved domain knowledge. This prompt is then passed to the LLM to generate an executable solution tailored to the specific CI/CD failure context.

\textit{\textbf{Automated Tool Selection and Execution:}} Once a solution is generated, LogSage leverages LLM tool-calling guided by the generated solution, the available tool set, and a specially designed prompt template. The LLM then selects and invokes the most appropriate tools from the internal automation toolkit. Currently, these tools include:

\begin{itemize}
    \item \texttt{lint\_fix}: automatically resolves lint-related issues without human intervention by leveraging the LLM’s contextual understanding and code generation capabilities;
    \item \texttt{image\_fix}: addresses container image version mismatches and automatically retrieves the correct images;
    \item \texttt{clean\_cache}: clears build or dependency caches;
    \item \texttt{apply\_resource}: requests access to internal resources.
\end{itemize}

After selecting the appropriate tool, the model fills in the required parameters and returns an interactive tool card with an execution button to the user interface. Upon user confirmation, the system invokes the selected tool and automatically re-triggers the CI/CD pipeline. If the remediation succeeds, the pipeline proceeds normally.

\begin{figure}[htbp]
    \centering
\begin{tcolorbox}[title=\textbf{Solution Generation Prompt Template}, colback=gray!5!white, colframe=gray!75!black, fonttitle=\bfseries, boxrule=0.5pt, arc=2pt, left=1pt, right=1pt, top=4pt, bottom=4pt]
    \scriptsize
    
    \textbf{\# Role:} You are a CI/CD pipeline failures fix assistant. Your task is to generate repair suggestions based on error logs, RCA report, and retrieved troubleshooting documents. \\
    
    \textbf{\# Skills:}
    \begin{itemize}
        \item \textbf{Targeted Suggestions:} Use RCA report and relevant documents to propose solutions. Each solution must align with the diagnosed error.
        \item \textbf{Multiple Options:} For each issue, list all applicable solutions. Choose document content most relevant to the CI/CD context.
        \item \textbf{Cite Source:} Every solution must mention the source document and include its official link (if provided). Do not fabricate URLs.
        \item \textbf{Command-level Guidance:} Provide actionable suggestions (e.g., install commands or CI config changes), not vague descriptions.
    \end{itemize}
    
    \textbf{\# Output Format:}
    \begin{itemize}
        \item Use section: \texttt{\#\#\# Solutions}.
        \item For a single root cause:
    \begin{verbatim}
    ### Solutions
    #### Problem: <summary>
    ##### 1. <Suggestion>
    ... Refer to [Doc Name](file_0)
    \end{verbatim}
        \item For multiple root causes:
    \begin{verbatim}
    ### Solutions
    #### Problem 1: ...
    ##### 1. ...
    ##### 2. ...
    #### Problem 2: ...
    \end{verbatim}
        \item Always include test name, file path, or dependency name when available.
    \end{itemize}
    
    \textbf{\# Notes:}
    \begin{itemize}
        \item Be concise and avoid redundant descriptions.
        \item Do not suggest solutions not backed by retrieved documents.
        \item Do not add file links; only use document links if provided.
    \end{itemize}
    
    \textbf{\# Constraints:}
    \begin{itemize}
        \item DO NOT generate new URLs — only use those in the documents.
        \item DO NOT omit document attribution in solutions.
        \item DO NOT provide vague advice like “check the config”.
    \end{itemize}
    
\end{tcolorbox}
\caption{Prompt template for LogSage’s solution generation stage.}
\label{suggestion_prompt}
\end{figure}

\section{Experimental Evaluation for RCA Stage}\label{sec:experimental}
In our survey of related work, we found that existing research provides suitable baselines for comparison with LogSage’s first stage (LLM-based RCA), but no appropriate counterparts exist for the second stage. Due to this limitation, we designed our experiments in two parts: for the first stage, we collected CI/CD task data from public GitHub repositories to compare LogSage with existing LLM-based RCA approaches, highlighting its theoretical performance advantages. The evaluation of the second stage, as well as the end-to-end effectiveness of LogSage, is presented in the following section \ref{sec:industrial} using data from ByteDance's real-world industrial deployments.

The research objectives for RCA in this section are:
\begin{itemize}
    \item \textbf{RQ1}: How does LogSage perform in root cause analysis precision compared to existing LLM-based baselines?
    \item \textbf{RQ2}: How does LogSage's cost efficiency in the root cause analysis stage compare to LLM-based baselines?
\end{itemize}
We conducted the following experiments around the above questions.

\subsection{Experimental Setup}
To comprehensively validate the performance of LogSage across various CI/CD scenarios, we built a public dataset for comparative evaluation against other LLM-based root cause analysis baselines. The models selected for this experiment are GPT-4o, Claude-3.7-Sonnet and Deepseek V3, with hyperparameters set to temperature = 0.1, and all other settings using the default values of the respective models.

\vspace{0.5em}
\subsubsection{\textbf{Dataset Description}}
To obtain representative and diverse CI/CD failure log cases, referring to \cite{Brandt2020}, we crawled the top 1,000 GitHub repositories by star count and filtered non-engineering-related repositories to ensure the cases have practical software engineering relevance. We then used the GitHub Action public API to retrieve the latest 300 CI/CD run logs for each qualifying repository and identified success and failure run pairs with the same workflow ID. After filtering out cases where no suitable run pairs exist, we obtained 367 cases from 76 repositories, each manually analyzed to identify the key failing log lines and the root causes. The first 117 cases were used to train the \textit{Drain} algorithm and build the few-shot log lines pool, with the remaining 250 cases used as the test set. Data collection was completed on April 25, 2025, and the dataset will be open-sourced on GitHub (see Appendix \ref{appendix}).

\vspace{0.5em}
\subsubsection{\textbf{Baseline Setup}}
According to our survey in Section \ref{sec:relatedwork}, LogSage is the first method specifically designed for CI/CD root cause analysis, and thus there are no existing baselines that can be directly compared. As the best available alternatives, we adopt LogPrompt~\cite{interpreteable} and LogGPT~\cite{loggpt} as baselines, since they represent recent LLM-based approaches for general log analysis and anomaly detection. We do not compare against non-LLM methods, as such approaches are typically restricted to specific log formats or require training models from scratch, which contrasts with LogSage’s training-free, plug-and-play integration into arbitrary CI/CD systems.

\vspace{0.5em}

\begin{itemize}
    \item \textbf{LogPrompt}: LogPrompt is an interpretable log analysis framework that utilizes prompt engineering techniques to guide LLMs in detecting anomalies without requiring any fine-tuning. It is designed for real-world online log analysis scenarios and emphasizes interpretability by generating natural language explanations alongside detection results. Evaluations across nine industrial log datasets show that LogPrompt outperforms traditional deep learning methods in zero-shot settings.
    \item \textbf{LogGPT}: LogGPT explores ChatGPT for log-based anomaly detection using a structured prompt-response architecture. It integrates chain-of-thought reasoning and domain knowledge injection to improve detection accuracy. The system generates structured diagnostic outputs with explanations and remediation suggestions. Experiments on benchmark datasets such as BGL demonstrate its competitiveness against state-of-the-art deep learning baselines under few-shot and zero-shot conditions.
\end{itemize}

\vspace{0.5em}

We evaluate LogPrompt and LogGPT using their original prompt templates, with optimal settings: LogPrompt (few-shot = 20, window = 100) and LogGPT (few-shot = 5, window = 30). As shown in Table~\ref{baseline}, LogSage is the only method that supports the full pipeline from interpretable root cause analysis to automated solution execution. LogPrompt can produce interpretable diagnostic outputs but lacks the ability to generate solutions. Although LogGPT claims to generate remediation suggestions, it operates without any external knowledge integration and relies solely on the LLM’s pre-trained knowledge, often resulting in hallucinated or impractical outputs.

\begin{table}[ht]
\centering
\caption{Comparison of LogSage and Baseline Methods}
\label{baseline}
\resizebox{0.95\columnwidth}{!}{
\begin{tabular}{@{}lccccc@{}}
\toprule
\textbf{Method} & \textbf{Anomaly} & \textbf{Interpretable} & \textbf{Solution} & \textbf{File-level} & \textbf{Automated} \\
 & \textbf{Detection} & \textbf{Root Cause} & \textbf{Generation} & \textbf{Processing} & \textbf{Execution} \\
\midrule
LogSage      & \checkmark & \checkmark & \checkmark & \checkmark & \checkmark \\
LogGPT       & \checkmark & \checkmark & $\circ$    & $\times$   & $\times$   \\
LogPrompt    & \checkmark & \checkmark & $\times$   & $\times$   & $\times$   \\
\bottomrule
\end{tabular}
}
\end{table}

\vspace{0.5em}
\subsubsection{\textbf{Metrics Description}}
We evaluate RCA task using standard metrics: Precision, Recall and F1-Score, with TP, FP, FN and TN defined specifically for this scenario as follows:

\begin{itemize}
\item \textbf{TP}: A correct detection. For LogSage, critical error log lines must overlap $\geq$ 90\% with ground truth; for LogPrompt and LogGPT, they must fall within a predefined context window.
\item \textbf{FP}: Detected lines that fail to meet the above criteria.
\item \textbf{FN}: LogSage produces no output; LogPrompt/LogGPT fail to detect anomalies despite overlap.
\item \textbf{TN}: Only applicable to LogPrompt and LogGPT when no detection intersects the context window.
\end{itemize}

\begin{table*}[ht]
\renewcommand{\arraystretch}{1.2}
\vspace{-0.5em}
\centering
\caption{Comparison of methods performance (Precision, Recall, F1-score) across different LLMs. \textsuperscript{*} indicates F1 scores that are statistically significant at $p < 0.001$ compared with all baselines (Wilcoxon signed-rank test).}
\label{tab:model_comparison}
\newcolumntype{M}[1]{>{\centering\arraybackslash}m{0.9cm}} 
\begin{tabular}{ll *{3}{M{0.9cm}} *{3}{M{0.9cm}} *{3}{M{0.9cm}}}
\toprule
\multirow{2}{*}{Group} & \multirow{2}{*}{Prompt Type} & \multicolumn{3}{c}{GPT-4o} & \multicolumn{3}{c}{Claude-3.7-Sonnet} & \multicolumn{3}{c}{Deepseek V3} \\
\cmidrule(lr){3-5} \cmidrule(lr){6-8} \cmidrule(lr){9-11}
 & & P & R & F1 & P & R & F1 & P & R & F1 \\
\midrule
LogSage   & N/A           & 0.9798 & \textbf{1.0000} & \textbf{0.9898}\textsuperscript{*} 
                          & 0.9837 & 0.9918 & \textbf{0.9878}\textsuperscript{*} 
                          & \textbf{0.9838} & 0.9959 & \textbf{0.9898}\textsuperscript{*} \\
\cmidrule(lr){1-11}
LogPrompt & CoT         & 0.8580 & 0.4654  & 0.6035 & 0.8467 & 0.3361 & 0.4812 & 0.8620 & 0.4477 & 0.5893 \\
          & InContext   & 0.6120 & 0.4530  & 0.5206 & 0.6864 & 0.4046 & 0.5091 & 0.6923 & 0.5018 & 0.5819 \\
\cmidrule(lr){1-11}
LogGPT    & Prompt-1    & 0.7280 & 0.3081  & 0.4330 & 0.8140 & 0.3075 & 0.4464 & 0.8006 & 0.3394 & 0.4767 \\
          & Prompt-2    & 0.7185 & 0.3484  & 0.4693 & 0.7715 & 0.3073 & 0.4396 & 0.7380 & 0.4438 & 0.5543 \\
\bottomrule
\end{tabular}
\vspace{-1em}
\end{table*}

\subsection{\textbf{RQ1}: Root Cause Analysis Precision Comparison}
To compare the RCA performance of LogSage with the two baseline methods and four prompt settings, we conducted experiments on three mainstream LLMs. The results are shown in Table \ref{tab:model_comparison}. We conducted Wilcoxon signed-rank tests comparing LogSage with each baseline variant, and all F1 score improvements are statistically significant ($p < 0.001$). 

It is evident that LogSage performs consistently well across all models, with near-perfect scores and minimal fluctuation, indicating high precision and robustness in handling diverse CI/CD log formats, lengths, and error types. In contrast, LogPrompt's performance suffers in precision, while LogGPT shows significant variability, with both methods exhibiting low recall, which suggests that their window-based sampling might miss key contextual information, leading to errors in anomaly detection.

\begin{tcolorbox}[
    enhanced,
    breakable,
    colback=gray!15,
    colframe=gray!50, 
    arc=4pt,
    boxrule=1pt,
    drop shadow={black!10!white,opacity=0.3}, 
    width=\columnwidth,
    enlarge left by=0pt,
    left=5pt, 
    right=5pt, 
    top=3pt, 
    bottom=3pt, 
]
\textbf{RQ1 Result}: LogSage significantly outperforms the baseline methods in root cause analysis, demonstrating high precision, recall, and F1-score across various models, ensuring stability and reliability.
\end{tcolorbox}

\subsection{\textbf{RQ2}: Root Cause Analysis Cost Comparison}
\begin{figure}[htbp]
    \centering
    \includegraphics[width=0.45\textwidth]{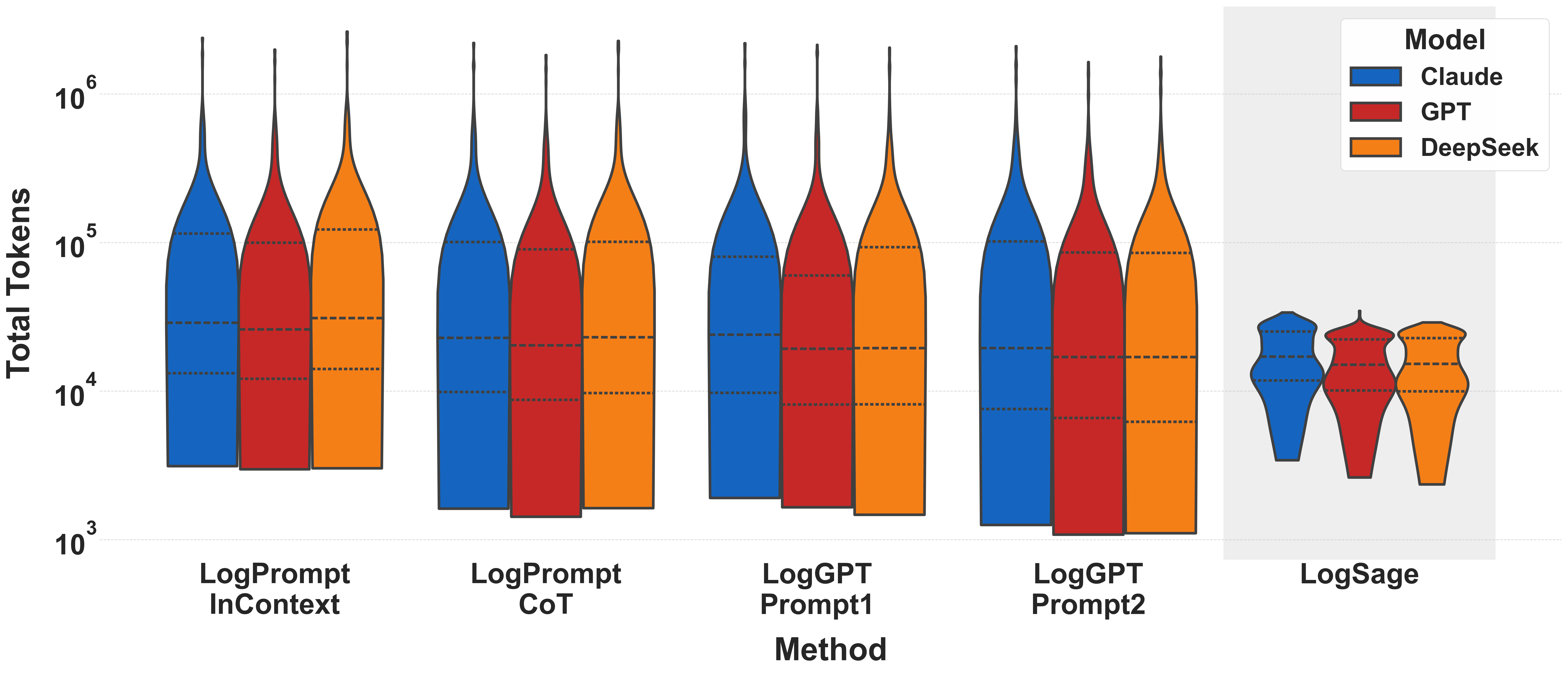}
    \caption{Token usage for RCA across methods and LLMs.}
    \label{violin_token}
\end{figure}

\begin{figure}[htbp]
    \centering
    \includegraphics[width=0.45\textwidth]{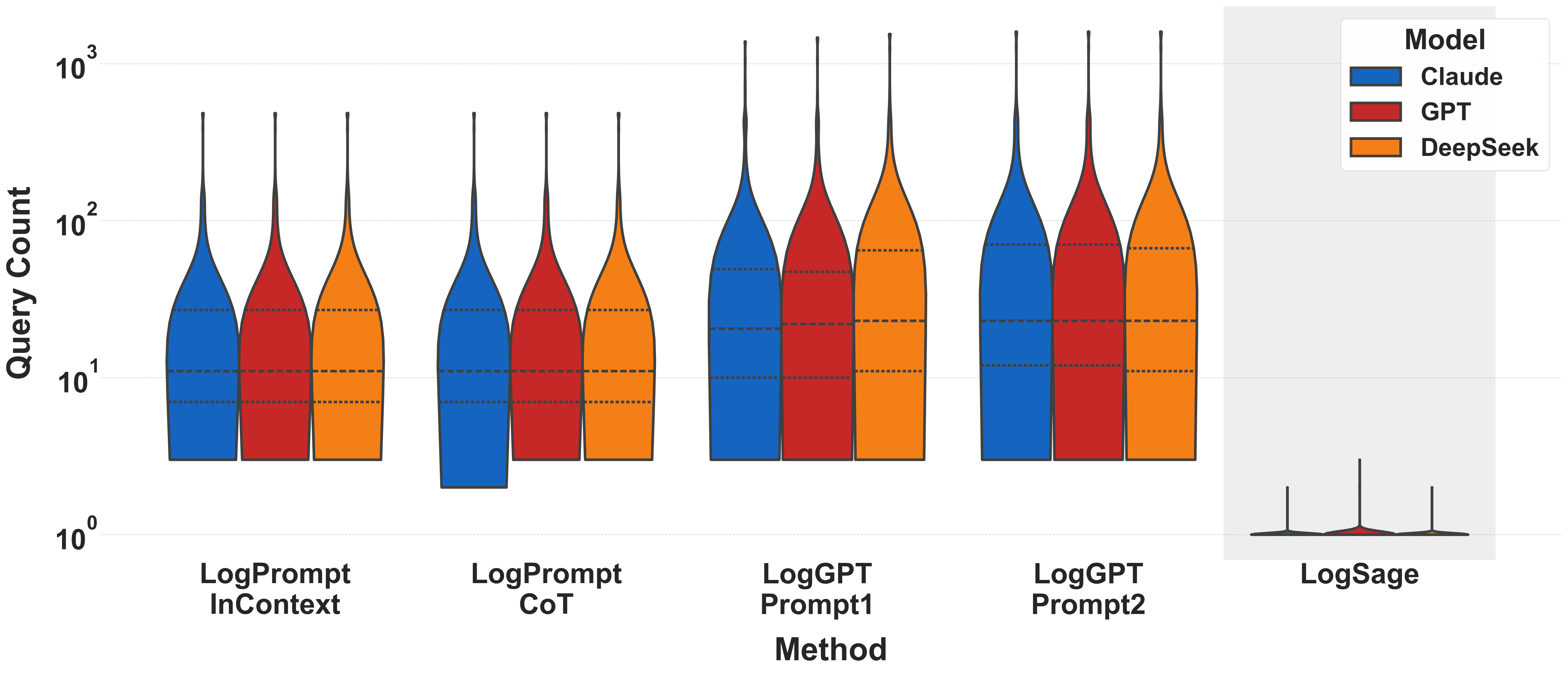}
    \caption{Query rounds for RCA across methods and LLMs.}
    \label{violin_query}
\end{figure}

Considering the cost and time consumption associated with LLMs, we analyzed the token usage and query rounds for each method from the baseline experiments. The distribution of the data is shown in the violin plots Figure \ref{violin_token} and \ref{violin_query}.

As shown in the token graph, the majority of CI/CD logs in the dataset have a length ranging from $10^4$ to $10^5$ tokens, with a long-tail distribution for some projects with unusually large logs. LogPrompt and LogGPT, which use stream-based processing through a sliding window without token length limits, show a rapid increase in token consumption as log length increases, leading to unnecessary cost in real-world scenarios. In contrast, LogSage not only achieves the best overall performance, but also benefits from a predefined token limit, maintaining stable token consumption across different models and ensuring more predictable costs.

As shown in the query round graph, LogPrompt and LogGPT exhibit a linear relationship between query rounds and token length. For most logs, they require nearly 10 query rounds to complete root cause analysis, with some extreme cases needing up to 1,000 rounds. Such a high number of query rounds is not only highly inefficient but also prone to failure in real-world application scenarios. In comparison, LogSage efficiently limits the query rounds to around 1 for most cases, with minimal retries in edge cases, offering significant time advantages.

\begin{table}[ht]
\centering
\caption{Efficiency Comparison Across Methods}
\label{tab:efficiency}
\resizebox{0.95\columnwidth}{!}{
\begin{tabular}{@{}lccc@{}}
\toprule
\textbf{Method} & \textbf{Avg. Tokens} & \textbf{Avg. Queries} & \textbf{Token Variability} \\
\midrule
LogSage    & \textbf{17{,}853}  & \textbf{1.0}   & \textbf{14.46\%} \\
LogPrompt  & 152{,}615 & 31.7  & 26.30\% \\
LogGPT     & 122{,}353 & 89.4  & 19.00\% \\
\bottomrule
\end{tabular}
}
\end{table}

We also analyzed the average token consumption and query rounds across the methods in Table \ref{tab:efficiency}. The average token consumption of LogPrompt and LogGPT are 152k and 122k tokens respectively, whereas LogSage maintains an average consumption of under 18k tokens—amounting to only 11.84\% of LogPrompt’s and 14.75\% of LogGPT’s. Moreover, LogSage demonstrates stable cross-model efficiency, with a normalized token variability of just 14.46\%, significantly lower than LogPrompt’s 26.30\% and LogGPT’s 19.00\%.

In terms of query rounds, LogSage exhibits strong practical applicability by requiring only 1.0 query round on average to perform accurate and actionable RCA for CI/CD failures across the entire test set. In contrast, LogPrompt, benefiting from a large window size, requires an average of 31.7 queries, while LogGPT, due to its smaller window size, incurs an average of 89.4 queries, rendering it nearly infeasible in real-world scenarios.

\begin{tcolorbox}[
    enhanced,
    breakable,
    colback=gray!15,
    colframe=gray!50, 
    arc=4pt,
    boxrule=1pt,
    drop shadow={black!10!white,opacity=0.3}, 
    width=\columnwidth,
    enlarge left by=0pt,
    left=5pt, 
    right=5pt, 
    top=3pt, 
    bottom=3pt, 
]
\textbf{RQ2 Result}: LogSage efficiently completes root cause analysis in an average of 1 query round with only 11. 84\% to 14. 75\% token consumption compared to baseline methods, making it a highly cost-effective solution in real-world production environments.
\end{tcolorbox}

\section{Industrial Deployment Validation}\label{sec:industrial}

We conducted a year-long online deployment and manual evaluation of LogSage to assess its accuracy and effectiveness in real-world CI/CD environments.

\vspace{0.5em}
\subsubsection{\textbf{Integration Method}}
\begin{figure}[htbp]
    \centering
    \includegraphics[width=0.45\textwidth]{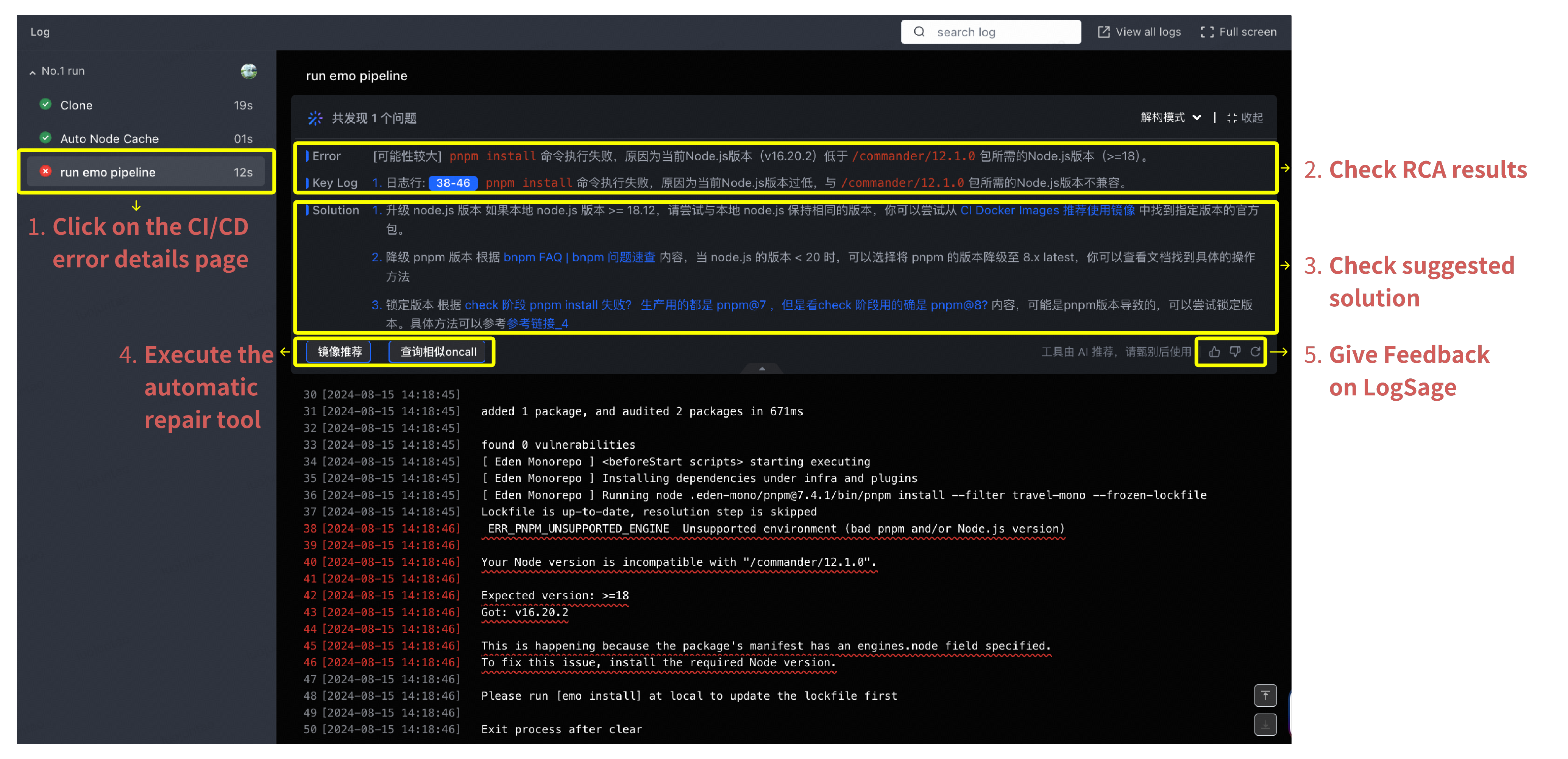}
    \caption{Screenshot of the online user interface.}
    \label{UI}
\end{figure}

LogSage was directly embedded into the company’s internal CI/CD platform (Fig.~\ref{UI}). When a CI/CD run failed, users could invoke LogSage from the failure page, review the diagnosed root cause and suggested fix, optionally trigger the repair tool, and provide feedback. Successful fixes automatically retriggered the pipeline, ensuring seamless integration and minimal workflow disruption.

\vspace{0.5em}
\subsubsection{\textbf{Deployment Scope}}
\begin{figure}[htbp]
    \centering
    \includegraphics[width=0.5\textwidth]{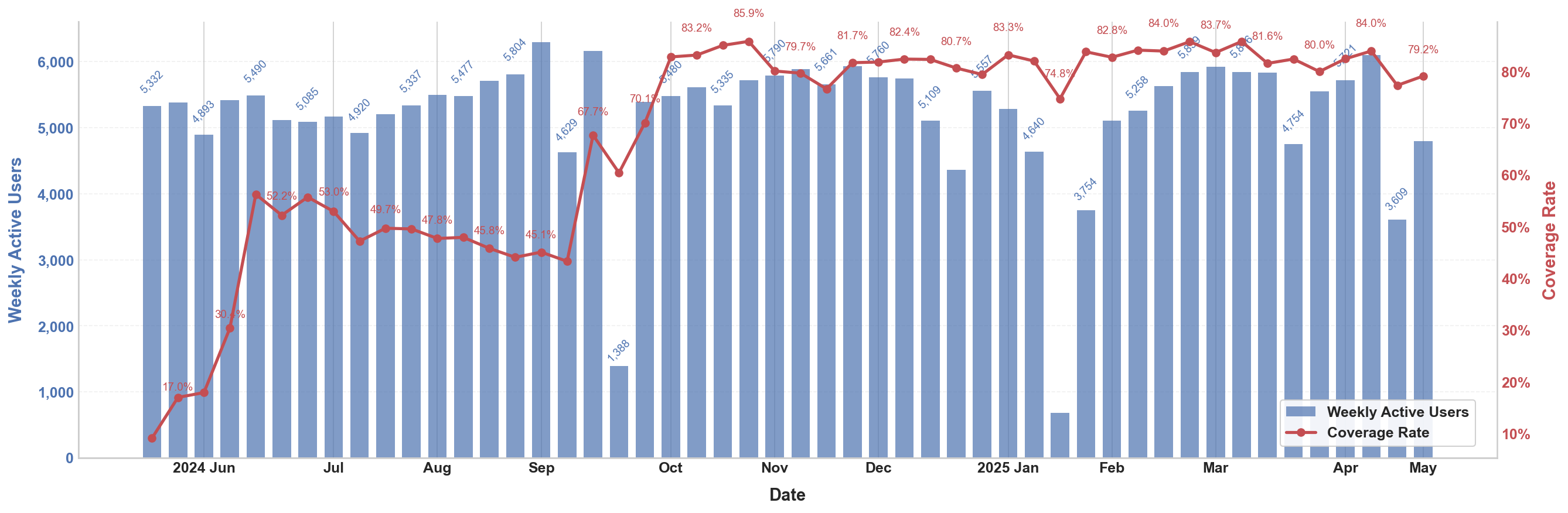}
    \caption{Weekly active user count and coverage rate.}
    \label{active_user}
\end{figure}

Since May 2024, LogSage has processed \textbf{1,070,613} CI/CD failures across the company, serving \textbf{36,845} developers, and has been made available to all R\&D teams using CI/CD services company-wide. Weekly active users exceeded \textbf{5,000} and coverage has stayed above \textbf{80\%} since Oct 2024 (Fig.~\ref{active_user}), showing broad and sustained adoption.

\vspace{0.5em}
\subsubsection{\textbf{Online Effectiveness}}
During the online deployment phase of LogSage, we evaluated its two-stage accuracy through manual assessment and measured the availability of its four automated remediation tools via automatic logging.

\textbf{Two-Stage Accuracy.}
The two-stage accuracy was evaluated based on weekly random sample of 150 online cases, which were manually reviewed and scored. For the RCA stage, experienced engineers examined the corresponding CI/CD task records and failure logs, manually debugging each case to determine the true root cause, which was then compared against LogSage’s output. The evaluation criteria for the solution generation stage are summarized in appendix \ref{appendix}, focusing primarily on the relevance of the generated solution to the actual root cause, the relevance of the retrieved knowledge, and the executability of the final recommendation.

\begin{figure}[htbp]
    \centering
    \includegraphics[width=0.45\textwidth]{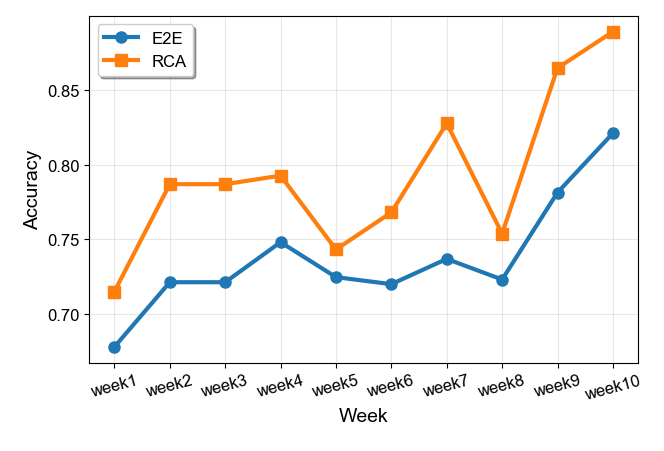}
    \caption{Human-Annotated Online Accuracy}
    \label{online_acc}
\end{figure}

The evaluation team independently assessed the outputs from Stage I and Stage II to measure the online accuracy of each stage. Due to the high cost of manual evaluation, assessments were conducted at fixed intervals only during the early rapid iteration phase of the project. The evaluation results from the 10 weeks' iteration period between August 2024 and October 2024 are shown in \ref{online_acc}. As shown in the figure, LogSage achieves over \textbf{85\%} RCA accuracy and over \textbf{80\%} end-to-end accuracy in real-world deployment scenarios, clearly demonstrating its usability in production environments.

\textbf{Auto-Repair Effectiveness.}
To assess the impact of auto-remediation, we collected engineering metrics on:

\begin{itemize}
\item Tool Coverage Rate: Ratio of cases where LogSage successfully recommended a repair tool over all cases that reached the solution generation stage.
\item Re-run Success Rate: Proportion of re-runs that passed automatically after executing the suggested fix.
\end{itemize}

These metrics were gathered via system logging. The average tool coverage rate was 22.4\%, with a median of 16.7\%, reflecting that overall tool support across CI/CD failures remains partial and that there is significant room to expand coverage. Nevertheless, when tools were applicable, LogSage achieved a re-run success rate of 47.4\% on average, with a maximum of 60.0\%, median of 47.1\%, and minimum of 36.8\%. These results indicate that nearly half of the supported failures could be successfully resolved without human intervention—significantly reducing developer effort and turnaround time.

Taken together, these findings underscore both the practical feasibility and long-term potential of integrating automated repair into CI/CD workflows. As the first large-scale industrial deployment of its kind, LogSage demonstrates not only immediate operational value but also a promising direction for further improvement.

\vspace{0.5em}
\subsubsection{\textbf{User Survey}}
Based on a recent survey of 191 front-line developers, LogSage received an average satisfaction score of 7.97 out of 10, reflecting strong acceptance and perceived utility in production workflows. Developers highlighted its ease of use, actionable suggestions, and seamless integration, while also providing feedback that has informed subsequent improvements.

\section{Conclusion \& Future Work}\label{sec:conclusion}
In this paper, we present LogSage, the first end-to-end LLM-based framework for CI/CD failure diagnosis and automated remediation, validated through large-scale industrial deployment. LogSage operates in two complementary phases. In the offline preparation phase, it leverages log template extraction and enterprise knowledge integration to construct reusable references for diagnosis. In the online operational phase, it performs root cause analysis by filtering, expanding, and pruning failed logs, followed by dynamically assembled diagnostic prompts and multi-route knowledge retrieval. Finally, LogSage generates executable solutions through LLM-guided tool selection and automated reruns, enabling accurate, interpretable root cause analysis and effective remediation of complex CI/CD failures.

Empirical evaluations across multiple LLM backends and baselines show that LogSage achieves significant improvements in both performance and efficiency. It outperforms state-of-the-art LLM-based methods in RCA precision while reducing token consumption by over 85\% compared to prior approaches. In industrial settings, LogSage demonstrates sustained adoption, processing over 1 million CI/CD failures and maintaining high user coverage with end-to-end precision exceeding 80\%.

While encouraging, several aspects warrant discussion. The effectiveness of the \textit{log diff} strategy relies on the repetitive nature of CI/CD pipelines: despite configuration heterogeneity, executions of the same pipeline remain largely stable across runs, allowing recurring outputs to be filtered regardless of system differences. Although our deployment used internal infrastructures such as Feishu documents and vector databases, the design is infrastructure-agnostic and can be instantiated with alternative knowledge bases in other organizations. Our baseline selection focused on LLM-based methods, as traditional CI/CD or AIOps approaches typically require training from scratch or depend on rigid log formats, making them less comparable to LogSage's training-free design. Finally, relying solely on LLMs for diagnosis introduces risks, as domain-specific semantics may not always be fully captured, leaving room for hallucinations or semantic gaps. These considerations highlight both the strengths and boundaries of LogSage, while pointing toward opportunities for broader integration.

In future work, we plan to upgrade LogSage into a more autonomous and adaptive LLM-Agent capable of orchestrating complex remediation workflows through iterative reasoning and proactive decision-making. We also aim to extend its scope beyond reactive failure handling to broader DevOps scenarios, including fault prediction, anomaly prevention, and automated incident response. These directions involve integrating with observability tools, modeling failure trends, and aligning with real-world operational workflows—pushing LogSage toward becoming an intelligent and proactive DevOps collaborator.

\appendices
\refstepcounter{section}\label{appendix}
\section*{APPENDIX \Alph{section}}

\textbf{Dataset Availability:} The dataset is publicly available at \url{https://github.com/ByteLuo1029/dataset}.

\vspace{0.5em}

\begin{tcolorbox}[
    title=\textbf{Manual Evaluation Criteria for Solutions},
    colback=gray!5!white,
    colframe=gray!60!black,
    fonttitle=\bfseries,
    boxrule=0.3pt,
    arc=1pt,
    left=3pt, right=3pt, top=2pt, bottom=2pt,
    before skip=2pt, after skip=2pt,
    enhanced,
    sharp corners,
    colbacktitle=gray!10!white,
    coltitle=black,
]
\scriptsize
\textbf{2 Points (E2E Good Case):}
\vspace{-0.3em}
\begin{itemize}
    \item \textbf{Non-code issues:} The solution provides \emph{direct and actionable instructions}, such as specific tools or recommended container images.
    \item \textbf{Code issues:} The solution clearly identifies the exact file and location of the issue based on logs, and proposes a concrete fix or a well-referenced example. 
\end{itemize}

\textbf{1 Point:}
\vspace{-0.3em}
\begin{itemize}
    \item \textbf{Non-code issues:} The solution is closely aligned with the root cause but only provides \emph{indirect suggestions} (e.g., vague instructions without actionable configuration or tooling).
    \item \textbf{Code issues:} The solution references the relevant logs and proposes a plausible fix, but \emph{fails to specify the exact file or code location}.
\end{itemize}

\textbf{0 Point:}
\vspace{-0.3em}
\begin{itemize}
    \item The solution is irrelevant to the true root cause, lacks sufficient log grounding, or includes incorrect content possibly due to hallucinated knowledge (e.g., fabricated repository names).
\end{itemize}

\textbf{Penalty:}
\vspace{-0.3em}
\begin{itemize}
    \item If any document link in the solution is invalid or broken, \emph{deduct 1 point} from the overall score.
\end{itemize}
\end{tcolorbox}

\vspace{-0.3em}

\end{document}